# Strain-induced interlayer magnetic coupling spike of two-dimensional van der Waals material $Fe_5GeTe_2$

Wen-Qiang Xie[1,2], Chang-Chun He[2], Xiao-Bao Yang[2], Yu-Jun Zhao[2,†], Wen-Tong Geng[1,†]

[1]School of Materials Science and Engineering, Hainan University, Haikou 570228, China

[2]Department of Physics, South China University of Technology, Guangzhou 510640, China

[†]Corresponding author: Yu-Jun Zhao (E-mail: zhaoyj@scut.edu.cn.) Wen-Tong Geng (E-mail: geng@hainanu.edu.cn)

## Abstract:

A stronger interlayer magnetic coupling (ILMC) can open up new opportunities in spintronics devices for $Fe_5GeTe_2$ (F5GT), a demonstrated two-dimensional (2D) van der Waals (vdW) material with high Currie temperature. Here we observe an extraordinary ILMC spike in F5GT, jumping from 1.15 to 12.79 meV/f.u, by applying a 3% in-plane strain. This spike is mainly ascribed to a significant increase in the magnetic moment of the Fe5 ion. Moreover, the applied in-plane strain can also significantly enhance the magnetic anisotropy energy (MAE) of the system, triggering the transition between the in/off-plane configurations in multi-layer F5GT.

## 1. Introduction

Magnetism in two-dimensional (2D) materials has attracted enormous attention in technological advances, as its potential application ranges from topological magnonics to low-power spintronics and quantum computing [1]. One of the promising features of 2D magnetism is the ability to fabricate various 2D heterostructures with engineered strains, chemistry, and electrical properties [2]. Many spintronic devices are constructed by several individual 2D layers bonding with the weak van der Waals (vdW) interaction, which has empowered these materials with the capacity for more profound applications, such as giant magnetoresistance (GMR) and tunneling magnetoresistance (TMR) [3-6]. Magnetic vdW materials are expected to open up a wide range of possibilities for spintronic devices.

Among the families of 2D vdW material, $Fe_5GeTe_2$ (F5GT) is a newly synthesized and promising vdW ferromagnet for spintronic devices, due to its high Curie temperature (about 293 K) [7] and novel spintronic properties [8, 9]. Strain engineering is an exciting and feasible method to manipulate the properties of materials in both theoretical and experimental investigations, which creates various opportunities for the study of new fundamental physics and applications of 2D vdW materials [10-13].

Meanwhile, 2D vdW materials were often transferred or grown onto a substrate such that the strain can be introduced by the substrate [10]. It has been observed that the magnetic properties of vdW materials are sensitive to applied strain [14-17].

Theoretical investigation of the magnetic and electronic properties variation of strained single-layer F5GT has been reported [18]. However, spintronic devices are mainly based on multilayer vdW materials or heterostructures, such as GMR and TMR devices. The presence of interlayer coupling could dramatically change the performance of devices[19, 20]. The interlayer coupling can be divided into two parts, i.e., the interlayer chemical coupling (ILCC) and interlayer magnetic coupling (ILMC) [21]. Earlier reports also illustrated the significance of ILMC in terms of magnetic properties [9, 22]. Therefore, investigating the strain influences on multi-layer F5GT is crucial for the design of spintronic devices.

Here, we report an impressive ILMC spike induced by in-plane strain in vdW F5GT. This extraordinary effect has significant implications for the design and performance of spintronic devices. In addition, we provide insight into the mechanism behind this ILMC spike through electronic structure analysis.

## 2. Computational details

The calculations were conducted based on the density functional theory [23, 24], by using the Vienna *ab initio* simulation package (VASP) code [25]. The projector augmented-wave (PAW) [26] method was employed with the Perdew-Burke-Ernzerhof (PBE) type generalized gradient approximation (GGA) for describing the exchange-correlation potential [27]. A cutoff energy of 560 eV for the plane wave basis set was applied to ensure an energy convergence of $10^{-6}$ eV for the electronic structure. The reciprocal space integrations were sampled with a *k*-mesh of 10×10×1 obtained by the Gamma-centered Monkhorst-Pack method [28]. For the PBE functional, the vdW effect was considered by the DFT-D3 method with Becke-Jonson [29]. The interlayer magnetic coupling (ILMC) and the interlayer chemical coupling (ILCC) are defined as follows:

$$\text{ILMC} = E_{\text{FM}} - E_{\text{AFM}}. \tag{1}$$

$$\text{ILCC} = E_{\text{multi-layer}} - E_{\text{single-layer}}. \tag{2}$$

Here, $E_{\text{AFM}}$ is the energy of anti-ferromagnetic (AFM) Fe$_5$GeTe$_2$ (F5GT) and $E_{\text{FM}}$ the ferromagnetic (FM) F5GT. $E_{\text{multi-layer}}$ and $E_{\text{single-layer}}$ indicate the energy of multi-layer (Bulk and double layers) and single-layer F5GT. Magnetic anisotropy energy (MAE) is defined as:

$$\text{MAE} = E_{\text{in-plane}} - E_{\text{off-plane}}. \tag{3}$$

Here, $E_{\text{in-plane}}$ stands for the energy with in-plane magnetization (parallel to F5GT plane), while $E_{\text{off-plane}}$ for the off-plane magnetization (perpendicular to F5GT plane). To ensure the accuracy of MAE, the *k*-mesh was set to 12×12×1. The MAE changes within 0.04 meV when *k*-mesh increases from 12×12×1 to 16×16×1. The magnetic interaction was studied using OpenMX [30], a density functional theory-based code, and a post-processing code for OpenMX based on Green's function [31]. We have conducted the related calculation using the PBE functional, with a *k*-mesh of 10×10×1 and a cutoff energy of 600 Ry. The criterion of energy convergence OpenMX is set to $1.0 \times 10^{-6}$ Hartree/Bohr.

## 3. Results and discussion

**3.1 The interlayer magnetic coupling spike induced by in-plane strain**

The crystal structure of $Fe_5GeTe_2$ (F5GT) is depicted in Figure 1 (a). Our model uses data collected on quenched crystals with the space group of *R3m*, (No. 160) [32]. The models adopted in our calculation are shown in Figure 1 (b). As described in our previous study [8], the lattice parameters calculated by PBE and LDA+U functionals are well in line with the measured values [32]. The deviation is only -0.07 Å (PBE) and 0.03 Å (LDA+U), respectively. The PBE functional underestimates the magnetic moments by 0.32 μB/Fe ion; by comparison, the LDA+U functional overestimates by 1.04 μB/Fe ion. Here, we choose PBE functional for further calculations. Notably, the PBE calculated energy of the AFM configuration is lower than the FM configuration by 1.45 meV per formula unit (meV/f.u.), implying that F5GT prefers an AFM coupling. Similar results were also discussed in Fe$_3$GeTe$_2$ system [33].

To our surprise, the evolution of ILMC (Figure 1 (c)) with in-plane strain exhibits an increase by an order, from 1.15 to 12.79 meV/f.u. at a 3% of in-plane strain. As the strain further increases, ILMC drops down rapidly, forming an ILMC spike at a 3% in-plane tensile strain. Furthermore, we also observe a sign-reverse of the ILMC at a 8% in-plane tensile strain, indicating that the ILMC changes from AFM to FM state. As for the off-plane strain, the ILMC evolves much more gently than that of the in-plane strain. The ILMC reaches the maximum of 2.3 meV/f.u. at a 2% off-plane strain and hits the minimum of -0.1 meV/f.u. at a -8% off-plane compressive strain (Figure 1 (d)). Therefore, the ILMC could also be tuned from AFM to FM state by applying a -8% off-plane compressive strain. Figure 1 (e) presents the ILCC evolution with in-plane and off-plane strains. An ILCC jump (red line) could also be observed when the in-plane strain is applied, in line with the observation of Figure 1 (c). Conclusively, we find an unusual ILMC spike and tunable ILMC properties in F5GT. These properties may hold a promising prospect for designing novel spintronic devices.

## 3.2 The mechanism of ILMC and the magnetic moment of Fe5 ion

To elucidate the ILMC spikes induced by the in-plane strain, we present the evolution of the magnetic moment of Fe ions under various strains. As plotted in Figure 2 (a), the magnetic moments of Fe1-Fe4 ions increase monotonically with the strain. However, the magnetic moments of Fe5 ion undergo a drastic jump from -0.1 μB to 1.4 μB at about 3% tensile strain, highly consistent with the change of ILMC. Moreover, the obvious spin polarization of the Fe5 ion can be observed at a 3% tensile strain (Figure 2 (b), insert). The rapid decline of ILMC may relate to the noticeable increase (0.13 *e*) in the Bader charge of Fe5 ion within the 3~4% in-plane strain range, which rearranges the charge density distribution remarkably. Regarding the off-plane strain (Figure S1), There is no significant enhancement observed in the magnetic moments of the five Fe ions, as their evolution appears to be smooth. Therefore, we mainly focus on the in-plane strain.

In Figur 2 (c) and (d), we present the partial density of states (PDOS) before and after the magnetic moment increase of the Fe5 ion. Since F5GT belongs to the hexagonal crystal system, the Fe ion satisfies the $D_{3h}$ symmetry. Its orbitals split into three states, namely $A'_1$ ($d_{z^2}$)、$E'$ ($d_{x^2-y^2}/d_{xy}$) and $E''$ ($d_{yz}/d_{xz}$). As shown in Figure 2 (c) and (d), $A'_1$ state is fully occupied at 2% and 3% in-plane strains. Both $E'$ and $E''$ states are partially occupied and distribute evenly at a 2% in-plane strain, leading to a low-spin configuration. As for a 3% in-plane strain, $E'$ and $E''$ states show an apparent polarization, resulting in the high-spin configuration.

Decomposed Mulliken population analysis [34, 35] is presented in Figure 3. Although Figure 3 (a) and (b) show $A'_1$ ($d_{z^2}$) is the dominant contribution in both the spin up and down channels, the difference in its values between these two channels is relatively smaller than that of $E'$ ($d_{x^2-y^2}/d_{xy}$) and $E''$ ($d_{yz}/d_{xz}$) (Figure 3 (d)). Thus, the $E'$ ($d_{x^2-y^2}/d_{xy}$) and $E''$ ($d_{yz}/d_{xz}$) states are the major contributors to the magnetic moment of Fe5. Meanwhile, the sharp increase/decrease of all spin up/down channel orbitals within the 2~3% strain range (Figure 3 (a) and (b)) indicates the presence of spin splitting in the Fe5 ion. This result is supported by the energy band diagram (Figure S2 (b) and (c)), where the spin up/down component moves down/up around the Fermi level with the strain.

Figure 3 (c) shows the sum-up value of spin up and down channels. The sum of $E'$ ($d_{x^2-y^2}/d_{xy}$) orbitals present a distinct increase at 2~3% strains, whereas the $E''$ ($d_{yz}/d_{xz}$) orbitals exhibit a decrease. It implies that the transfer of charge from $E''$ ($d_{yz}/d_{xz}$) orbitals to $E'$ ($d_{x^2-y^2}/d_{xy}$) orbitals is a crucial factor contributing to the spike observed in the ILMC. This result can also be confirmed by occupied bands (Figure S2 (a)) and band diagrams (Figure S2 (b) and (c)). The number of occupied bands in spin up channel increase from 39 to 40 at a 2% in-plane strain (Figure S2 (a)), in agreement with the ILMC spike. With a 3% in-plane strain applied, the blue-dashed band in the spin-up channel shifts downward and intersects the Fermi level within the $Y_2 \sim$ Gamma $k$-point region. The band in the aforementioned $k$-point region is primarily composed of the $E'$($d_{x^2-y^2}/d_{xy}$) orbitals (Figure S3). This finding aligns well with the Mulliken population analysis.

We propose that the increase in magnetic moment of the Fe5 ion is the driving force behind the observed ILMC. Moreover, based on the Bader, Mulliken population and band diagram analysis, the significant change in magnetic moment is a result of two processes as plotted in Figure 3 (e), i.e., the charge transfer from $E''$ ($d_{yz}/d_{xz}$) orbital to $E'$ ($d_{x^2-y^2}/d_{xy}$) orbital, and the spin splitting of spin up/down channels.

**3.3 The drastic change in intralayer exchange parameters**

The exchange parameters are of significance in determining the magnetic properties and Curie temperature of the single-layer system [36, 37]. As plotted in Figure 1 (b), each Fe5 ion has its first, second, and third nearest neighbor Fe ions, corresponding to the first ($J_1$), second ($J_2$) and third ($J_3$) intralayer Heisenberg exchange parameters. The distances from each Fe5 to its three neighbors are 2.4, 2.8, and 3.7 Å, respectively. The calculated $J_1$, $J_2$, and $J_3$ parameters with in-plane strain are presented in Figure 4. The exchange interaction ($J$) around Fe5 ion experiences a dramatic change within the in-plane strain range of 2%~3%. Specifically, the dominant contribution ($J_1$) soar to 13.0 (3%) from 3.2 (2%) meV. $J_2$ also presents a sharp increase from negative (-1.3 meV) to positive (6.1 meV). Conversely, $J_3$ displays an opposite trend, declining from a positive value of 1.0 meV to a negative value of -3.5 meV. The sum-up value of $J_1$, $J_2$ and $J_3$ demonstrates a noticeable improvement within 2~3% extensile train range, suggesting a substantial enhancement of intralayer magnetic coupling.

### 3.4 The evolution of magnetic anisotropy energy with thickness and strain

The Magnetic anisotropy energy (MAE), one of the most important properties of 2D magnetic materials, is sensitive to strain and film thickness [38]. To examine the impact of various staggered stackings, Three stack models are shown in Figure S4. All of them show consistent evolving trends in both ILCC and ILMC (Figure S4 (a) and (b)). These two couplings both display drastic jumps within the 2~3% train range, similar to the bulk model. As a result, we have chosen the AB stack model randomly for further studies.

The evolution of MAE in F5GT with various thicknesses and in-plane strains are presented in Figure 5. The acute increase in the magnetic moment of Fe5 ion has a huge impact on MAE, as demonstrated by the abrupt rise in MAE in the single-layer, double-layer, and bulk systems within the 2-3% strain range. Regarding the strain-tunable magnetic anisotropy, F5GT displays a preference for an in-plane magnetic configuration in the absence of strain. However, two distinct switching points between in-plane and off-plane magnetic configurations can be observed in single-layer F5GT at around -6% and 8% strains (Figure 5 (a)). After the introduction of interlayer coupling, the magnetic moment of Fe ion varies from 0.1 to -0.3 μB, resulting in the slight delay of the magnetic tipping point (Figure S4 (c) and (d)). As for MAE, the inclusion of interlayer interaction shifts the switching point, previously observed at around a 8% strain in the single layer, backwards to the 2~3% strain range (Figure 5 (b) and (c)). The MAE differences (between 2% and 3% in-plane strains) are strongly enhanced from 0.47 to 0.85 meV/f.u. with thicknesses. Compared to the double-layers, the bulk system exhibits stronger interlayer coupling, leading to a notably enhanced stability of the off-plane configuration. This is supported by the almost doubled value of its MAE at a 3% in-plane strain. In conclusion, the interlayer coupling in the F5GT system significantly enhances its MAE, leading to a switch between in-plane and off-plane configurations within the 2~3% in-plane strain range.

## 4. Conclusion

Our theoretical investigation of the evolution of interlayer coupling in F5GT with strains has revealed a remarkable spike in the ILMC. Within the 2~3% in-plane strain range, the ILMC increases rapidly from 1.15 to 12.79 meV/f.u. The acute enhancement in the magnetic moment of the Fe5 ion, from -0.1 μB to 1.4 μB at a 3% in-plane strain, coincides well with the ILMC spike. Moreover, the electronic structure analysis reveals that the mechanism involves charge transfer from $E'$ ($d_{x^2-y^2}/d_{xy}$) to $E''$ ($d_{yz}/d_{xz}$), and spin splitting of Fe5 ion. Additionally, the applied 3% in-plane strain may also result in a noticeable enhancement of the intralayer magnetic coupling. Incorporating interlayer coupling into the system can markedly increase its MAE, which triggers the transition of the in-plane and off-plane configurations within a strain range of 2~3%. With these distinctive characteristics, F5GT is promising for future spintronic applications and designs.

# Figures

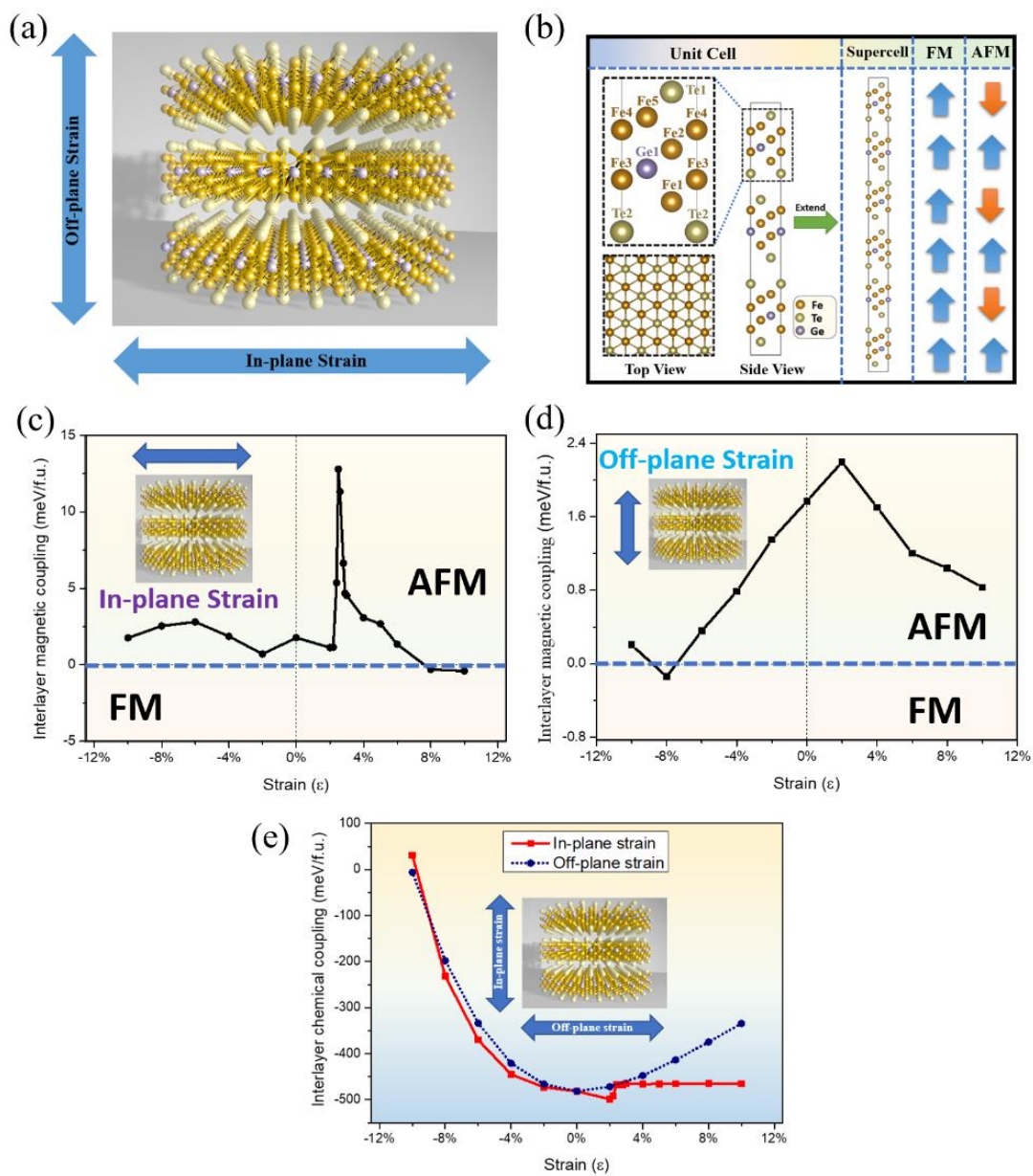

Figure 1: (a) The illustrative structure of Fe$_5$GeTe$_2$ and (b) the magnetic configurations of interlayer FM/AFM states. The interlayer magnetic coupling (ILMC) evolution with (c) in-plane and (d) off-plane strain. (e) The interlayer chemical coupling (ILCC) variation with in-plane and off-plane strain.

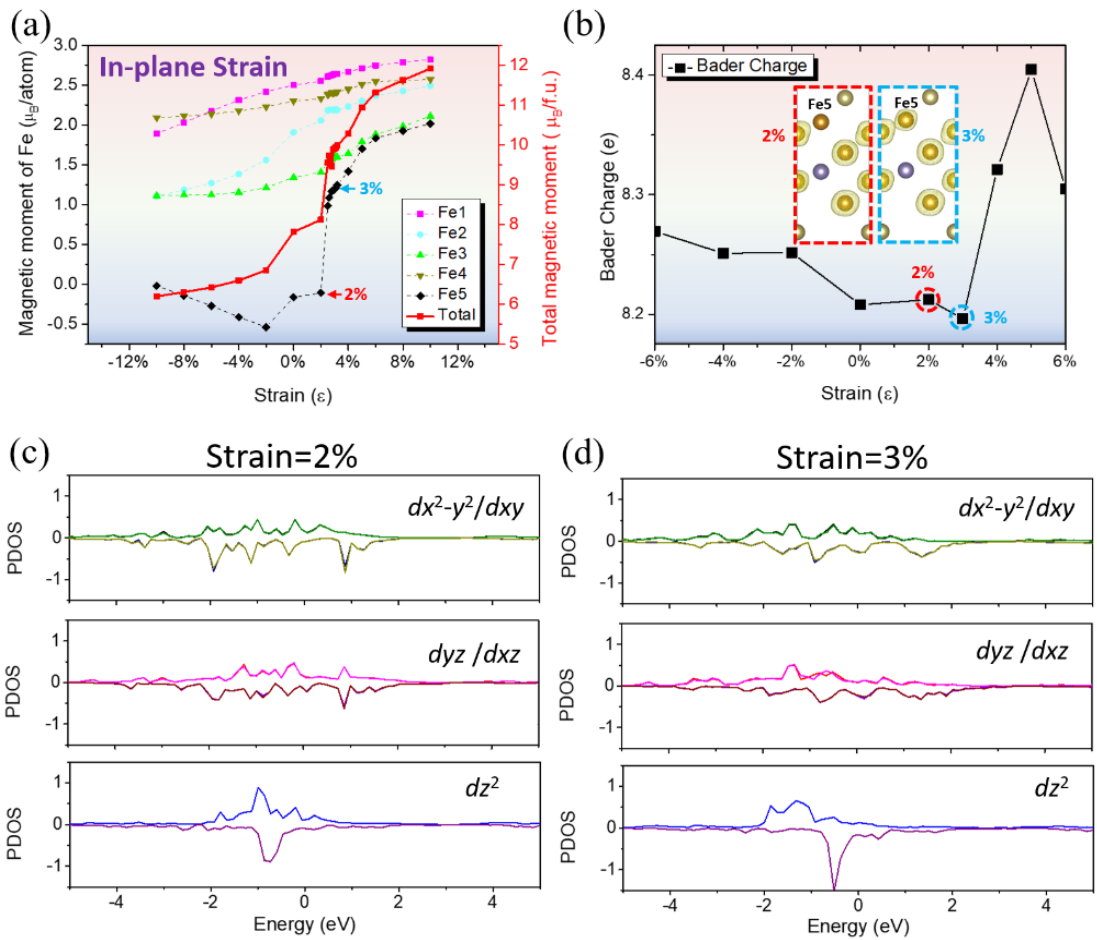

Figure 2: (a) The variation in the magnetic moments of Fe ions caused by the in-plane strain. (b) The Bader charge variation of Fe5 ion with in-plane strains. Insert: The spin density with 2% and 3% in-plane strains. Projected density of states (PDOS) of Fe5 ion under (c) a 2% and (d) a 3% in-plane strain.

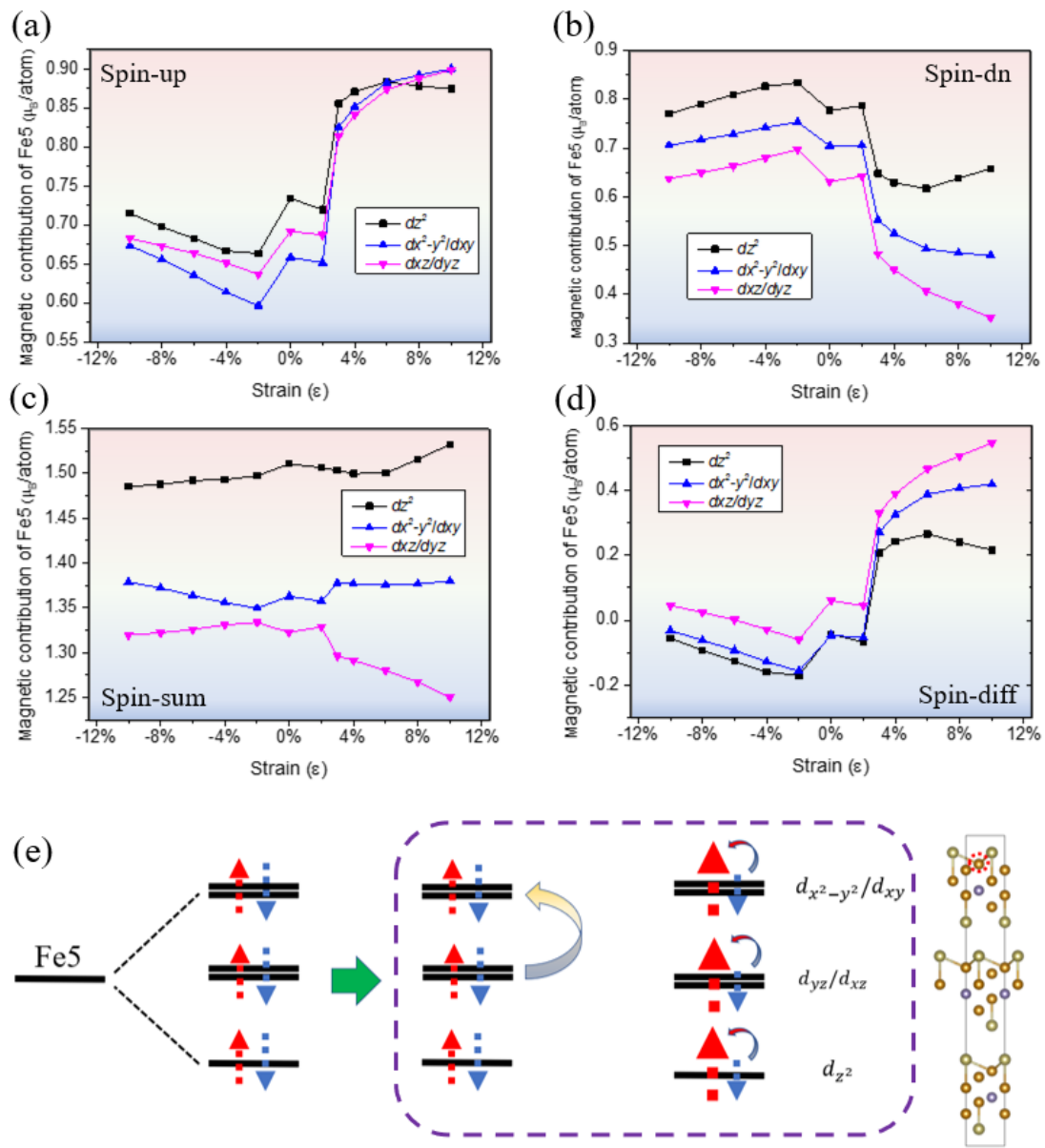

Figure 3: Mulliken population analysis of Fe5 ion: (a) decomposed spin up and (b) spin down channels. (c) The sum of spin up and down channels. (d) The difference between spin up and down channels. (e) The proposed mechanism of the acute increase of Fe5 magnetic moment.

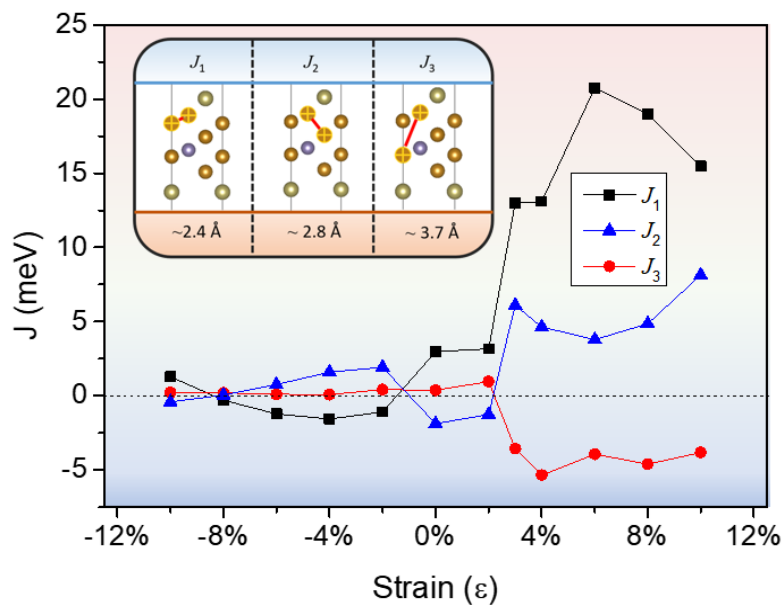

Figure 4: The evolution of the first ($J_1$), second ($J_2$) and third ($J_3$) nearest neighbor Heisenberg exchange parameters around the Fe5 ion.

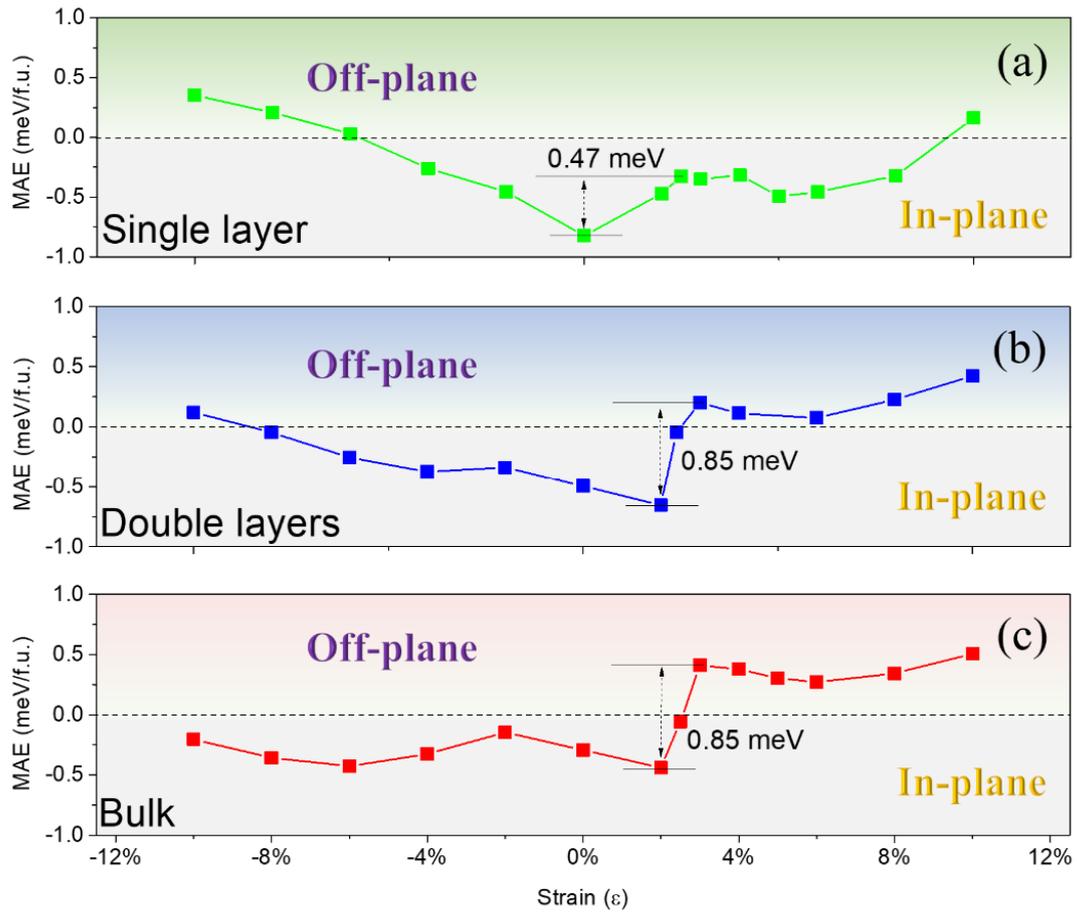

Figure 5: The calculated magnetic anisotropy energy (MAE) of (a) the single layer, (b) double layers, and (c) bulk models.

## Acknowledgements

This work was supported by the NSFC (Grant No. 12074126), the Foundation for Innovative Research Groups of the National Natural Science Foundation of China (Grant No. 51621001), and the Natural Science Foundation of Guangdong Province of China (Grant No. 2016A030312011). The computer times at the National Supercomputing Center in Guangzhou (NSCCGZ) are gratefully acknowledged.

# Supplementary Materials

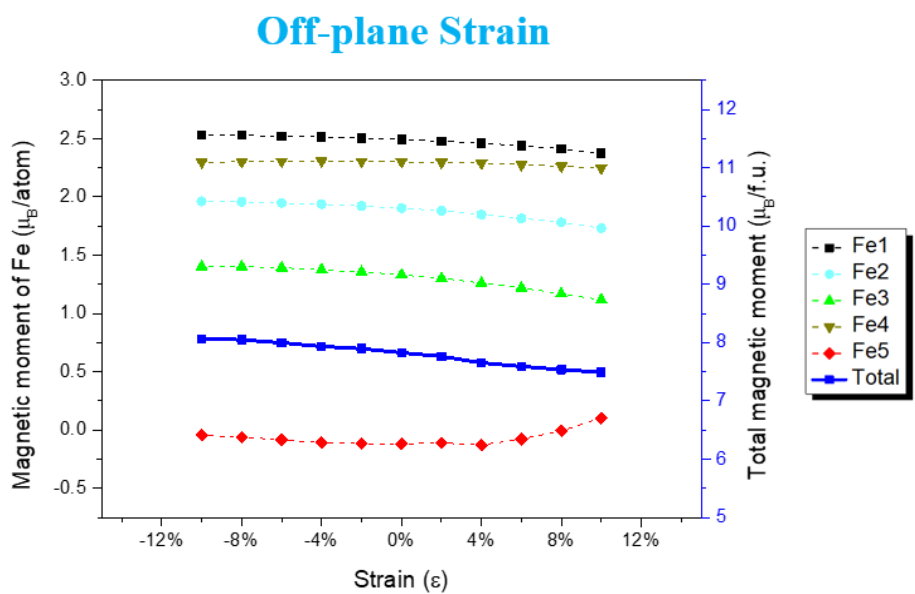

Figure S1: The variation in magnetic moments of Fe ions caused by the off-plane strains.

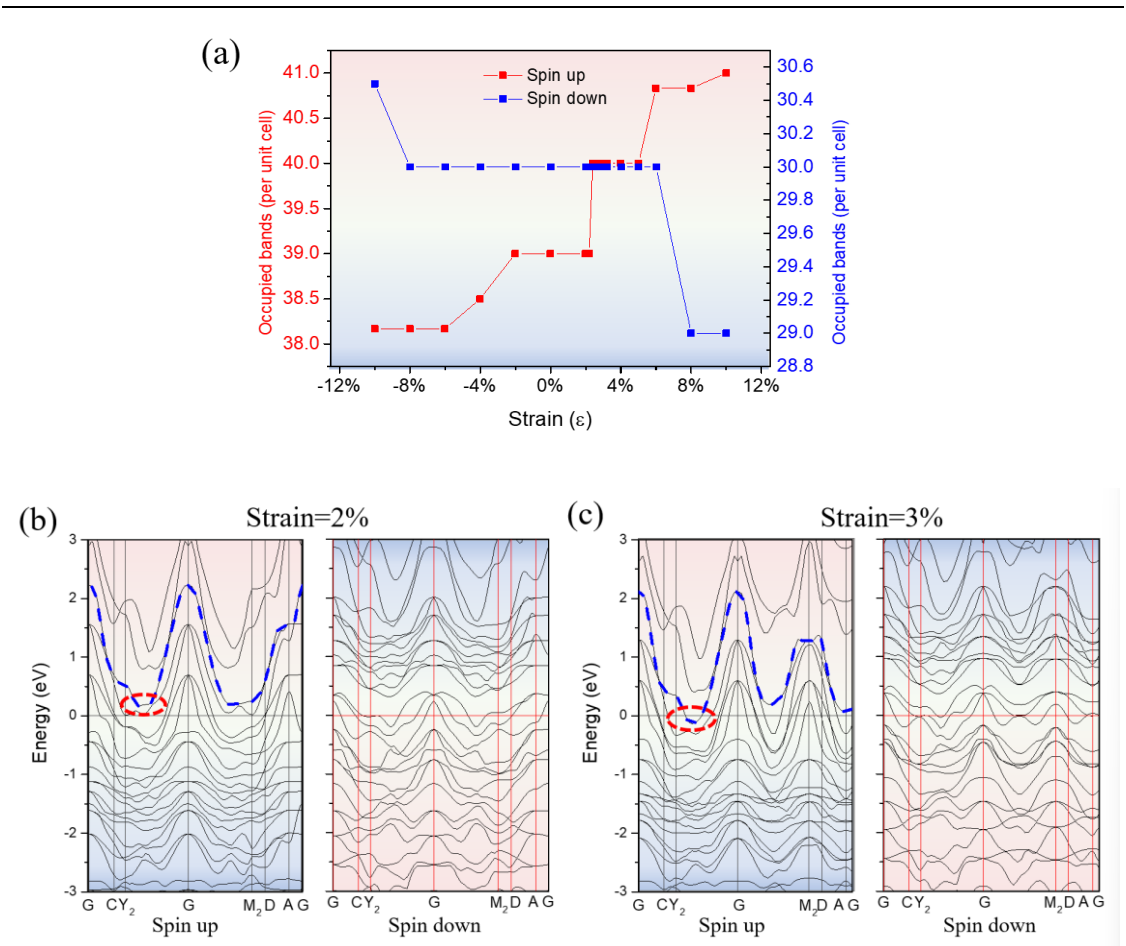

Figure S2: (a) The electron-occupied bands, a nearly vertical ascent of occupied bands from 39 to 40 at 2% in-plane strain. This ascent is highly in line with the ILMC spike and magnetic moment variation of the Fe5 ion. (b) Band structure diagrams of F5GT under (c) 2% and (d) 3% in-plane strain. Note that the occupied band may not be completely full.

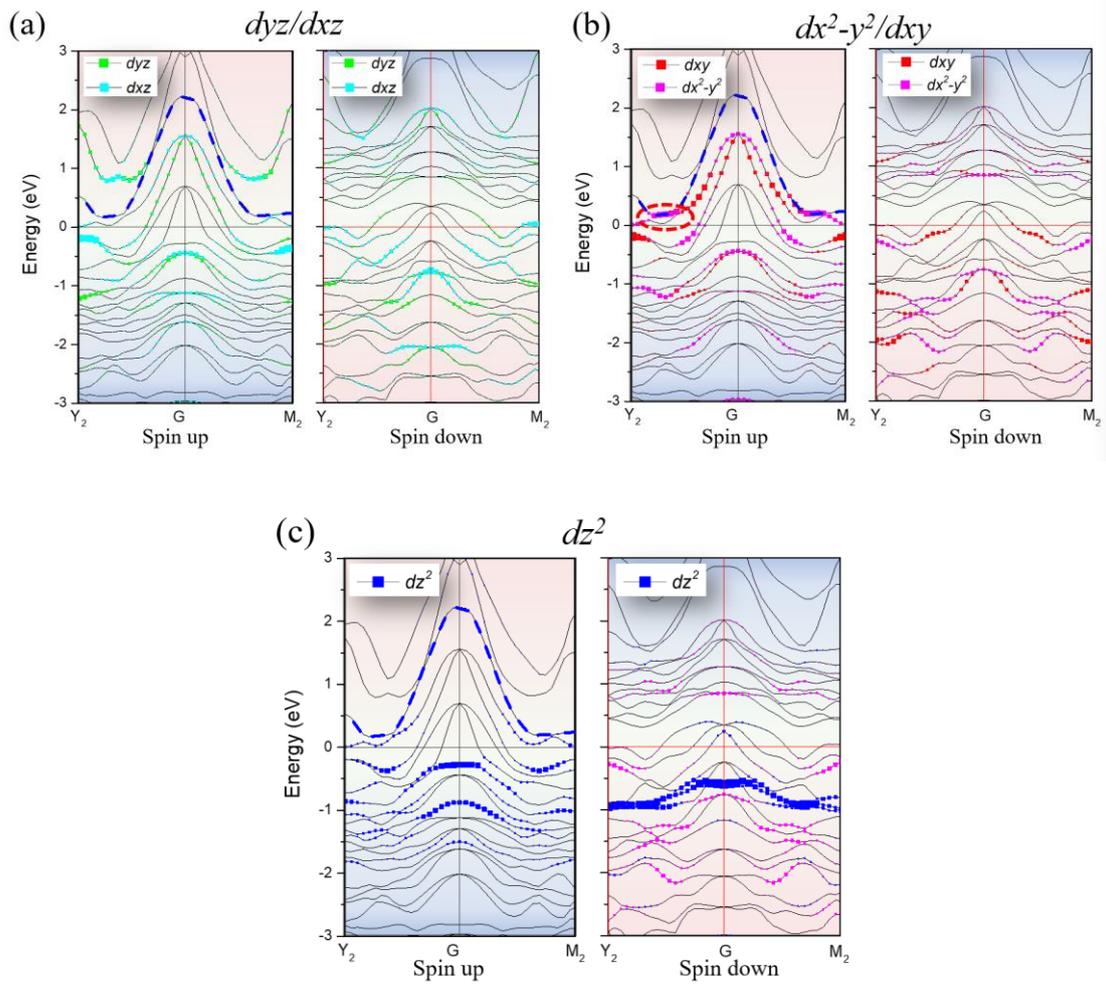

Figure S3: The projected band structures (Pband) of the Fe5 ion under 2% in-plane strain.

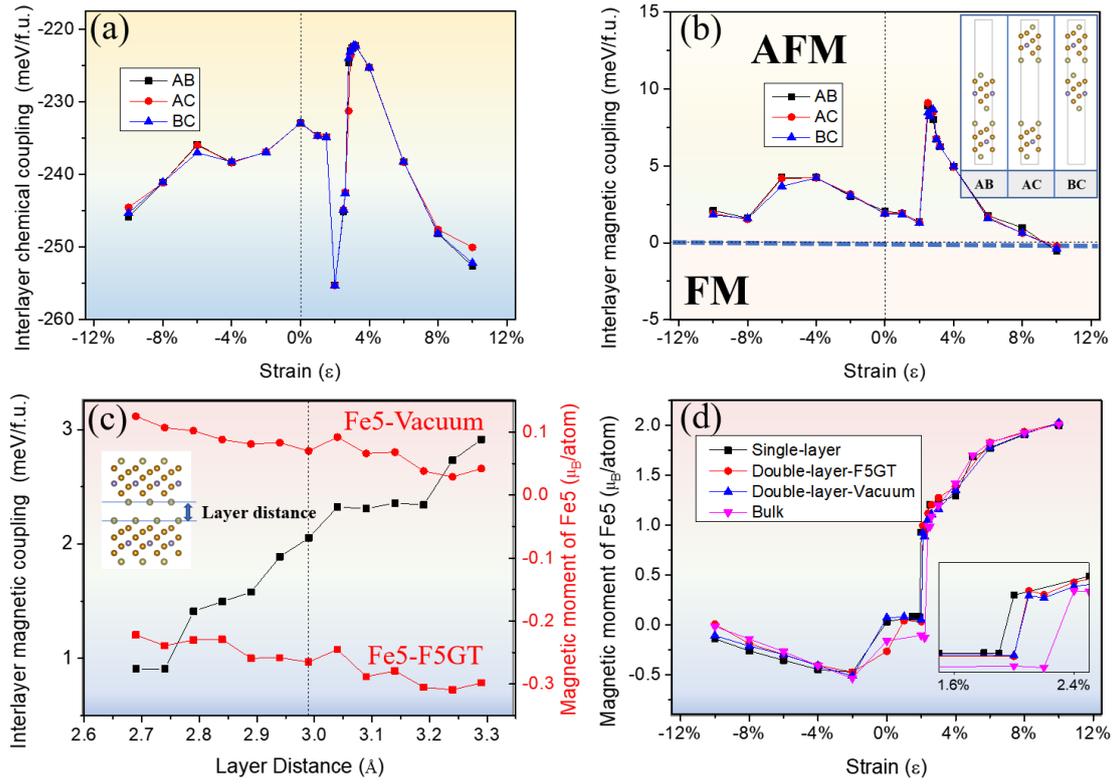

Figure S4: The variation of (a) ILCC and (b) ILMC of F5GT with three stacking models, insert: Computational models of AB, AC, and BC stackings; (c) The ILMC and magnetic moment of Fe5 ion evolves with the interlayer distance, the vertical dashed line represents the interlayer spacing without squeeze and stretch. The introduction of interlayer coupling can lead to a divengence of the magnetic moments of Fe5 ion, i.e. around 0.1 (vicinity to the vacuum, without interlayer coupling) and -0.3 µB (vicinity to the F5GT layer, with interlayer coupling), respectively; (d) Magnetic moments of Fe ions in single-layer, double-layer-vacuum (Fe5 ion vicinity to the vacuum layer), double-layer-F5GT (Fe5 ion vicinity to the upper F5GT layer) and bulk F5GT.